\documentclass[twocolumn,times]{aastex63} 

\usepackage{amssymb}
\usepackage{amsmath}
\usepackage{xspace}
\usepackage{epstopdf}
\usepackage[english]{babel}
\usepackage{mathrsfs}
\usepackage{mathtools}
\usepackage{graphicx}
\usepackage{listings}
\usepackage{xcolor}
\usepackage{array}
\usepackage{listings}

\graphicspath{{figures/png/}}

\shorttitle{Lithium \& Beryllium in NGC 752}
\shortauthors{Boesgaard et al.}

\begin{document}

\title{Lithium and Beryllium in NGC 752: An Open Cluster Twice the Age of the Hyades}

\author[0000-0002-8468-9532]{Ann Merchant Boesgaard}
\affiliation{Institute for Astronomy, University of Hawai`i, 2680 Woodlawn Drive, Honolulu, HI 96822, USA}

\correspondingauthor{Ann Merchant Boesgaard}
\email{annmb@hawaii.edu}

\author[0000-0001-7205-1593]{Michael G. Lum}
\affiliation{Institute for Astronomy, University of Hawai`i, 2680 Woodlawn Drive, Honolulu, HI 96822, USA}

\author[0000-0003-1125-2564]{Ashley Chontos}
\altaffiliation{NSF Graduate Research Fellow}
\affiliation{Institute for Astronomy, University of Hawai`i, 2680 Woodlawn Drive, Honolulu, HI 96822, USA}

\author[0000-0002-3854-050X]{Constantine P. Deliyannis}
\affiliation{Department of Astronomy, Indiana University, 727 East 3rd Street, Bloomington, IN 47405-7105, USA}

\begin{abstract}
\noindent The surface abundances of the light elements lithium (Li) and beryllium (Be) reveal information about the physical processes taking place in stellar interiors. The investigation of the amount of these two elements in stars in open clusters shows the effect of age on those mechanisms. We have obtained spectra of both Li and Be in main-sequence stars in NGC 752 at high spectral resolution and high signal-to-noise ratios with HIRES on the Keck I telescope. In order to make meaningful comparisons with other clusters, we have determined the stellar parameters on a common scale. We have found abundances of Li and Be by spectral synthesis techniques. NGC 752 is twice the age of the well-studied Hyades cluster. We find that 1) The Li dip centered near 6500 K is wider in NGC 752, having expanded toward cooler temperatures; 2) The Be dip is deeper in the older NGC 752; 3) The Li ``peak'' near 6200 K is lower by about 0.3 dex; 4) Although there is little Be depletion in the cooler stars, it is possible that Be may be lower in NGC 752 than in the Hyades; 5) The Li content in both clusters declines with decreasing temperature, but there is less Li in NGC 752 at a given temperature by $\sim$0.4 dex. These differences are consistent with the transport of the light-element nuclei below the surface convection zone as predicted by theory. That connection to rotational spin-down is indicated by the pattern of rotation with temperature in the two clusters.
\end{abstract}


\section{Introduction} \label{sec:intro}

Open clusters provide laboratories for studying many aspects of stellar structure and evolution because of the common origin, age, and composition of the stars in a given cluster. The changes with time are especially revealed by a comparison of properties in clusters of an array of ages. The surface abundances of Li, Be, and B in clusters enable exploration the otherwise unobservable interiors of stars. These elements reveal the internal structure of stars because they are easily destroyed inside stars by thermonuclear reaction, particularly (p,$\alpha$) type. The three elements are susceptible to destruction at different interior temperatures with Li being most fragile and B being more robust. The destruction of Li occurs above T = 2.5 $\times$ 10$^6$ K, Be above 3.5 $\times$ 10$^6$ K, and B above 5 $\times$ 10$^6$ K. Thus observations of the surface abundances reflect the depth and mechanisms of the mixing process(es) inside a star.

Cluster stars can show changes resulting from from stellar mass, age, evolutionary stage, composition, rotation and magnetism. Abundances of Li and Be can illuminate the effect of rotation and magnetic fields on stars and how stars evolve. When stars are formed they have angular momentum. A comprehensive approach to modelling stellar evolution with rotation was pioneered by \citet{endal1976,endal1981}, and includes effects of angular momentum loss from the surface and the action of a number of rotation-related instabilities in the interior. As stars evolve beyond the fully convective deuterium birthline along the Hayashi track, they contract and spin up. But a magnetically-driven wind causes the surface layers to spin down \citep{kawaler1987}, causing a gradient of rotation with depth to increase as stars age. In the outer layers, when this gradient becomes too large, a secular shear instability is triggered \citep{zahn1974} that requires redistribution of angular momentum and causes mixing \citep{pinsonneault1989}. As a result, the surface abundances of Li, Be, and B decline as mixing brings these elements to depths where they are destroyed \citep{chaboyer1995,sills2000,somers2016}. Since these elements survive to different depths, observations of two or more of them can provide very powerful means to evaluate the effects of physical mechanisms that depend on depth. For example, since the efficiency of the shear-induced mixing described above decreases with depth, rotational models incorporating this shear predict a very different ratio of surface Li to Be to B than other mechanisms that have been proposed to explain Li depletion. Such observations of two or more of these elements in the same stars that have strongly evidenced rotational mixing as the dominant cause of light element depletion in main sequence stars \citep{stephens1997,deliyannis1998,boesgaard1998,boesgaard2001,boesgaard2005,boesgaard2016}. For some recent reviews of these and related observational discriminators see \citet{cummings2017} and \citet{boesgaard2020}. With this study, NGC 752 joins studies of Li and Be in other key open clusters (e.g., \citealt{boesgaard2001}).

Lithium abundances have been studied in main sequence stars in NGC 752 first by \citet{hobbs1986} in 18 stars; later \citet{pilachowski1988} added six more. They confirmed F star Li dip that had been found in the Hyades by \citet{boesgaard1986}. \citet{pilachowski1988} commented that ``the dip may extend to slightly cooler temperatures'' in this older cluster than was found in the Hyades cluster.

In her study of Li in M 67, \citet{balachandran1995} reanalyzed the Li data from \citet{hobbs1986} in NGC 752 for comparison with M 67. In her work she used [Fe/H] = $-$0.15 for NGC 752 while more modern determinations have found values consistent with solar, e.g., \citet{carrera2011}: +0.08; \citet{maderak2013}: $-$0.06; \citet{blanco2014}: +0.04; \citet{lum2019}: $-$0.01.

A population census and membership analysis of NGC 752 has been done by \citet{daniel1994}. They distinguish cluster members from field stars by proper motions, radial velocities and photometry. Some more recent studies of NGC 752 are by \citet{maderak2013}, \citet{castro2016}, \citet{sestito2004}, \citet{agueros2018}, \citet{lum2019}. The papers by \citet{sestito2004} and \citet{castro2016} focus on Li.

The open cluster, NGC 752, has an important place in cluster studies because it is twice the age of the very well-studied Hyades cluster and about one-third the age of the Sun.

\section{Observations}

\begin{figure}
\centering
\includegraphics[width=\linewidth]{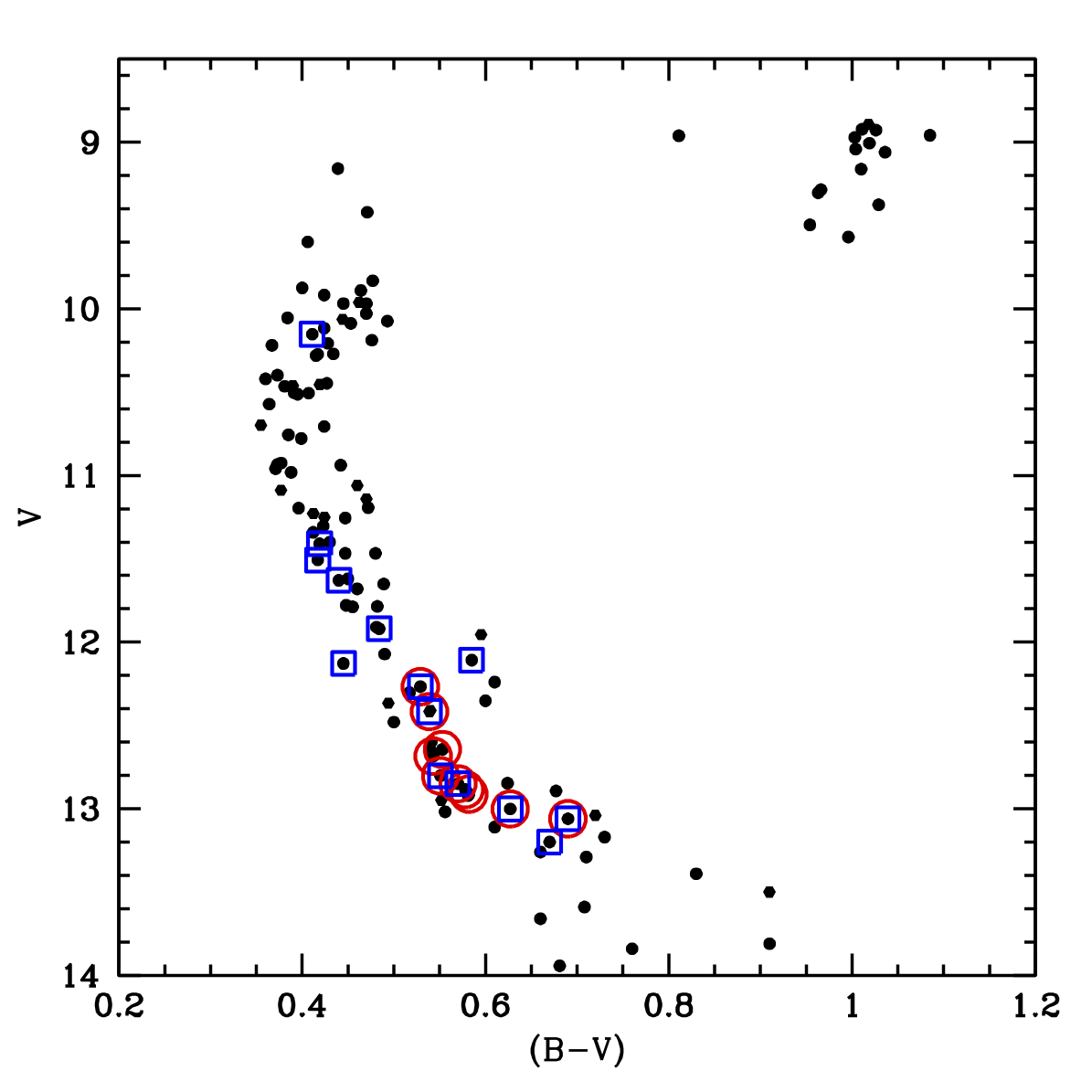}
\caption{Color-magnitude diagram for NGC 752 with data from \citet{daniel1994}. The 14 stars we observed for Be have blue squares and those 10 for Li have red circles. We observed both Li and Be in six stars.}
\label{fig:f1}
\end{figure}

We use the work of \citet{daniel1994} to make the color-magnitude diagram of NGC 752 shown in Figure \ref{fig:f1}. The stars we observed for Li are circled in red and those for Be are indicated by blue squares.

\subsection{Beryllium}

Observations were made of 14 dwarf stars in NGC 752 over four observing runs between 2014 and 2019 with HIRES \citep{vogt1994} on the Keck I telescope. The Be II resonance doublet is at 3130.42 and 3131.06 \AA {} close to the atmospheric cutoff. Atmospheric absorption and dispersion are strong in the ultraviolet region and so long exposure times are needed to obtain spectra with good signal-to-noise ratios. We tried to observe our stars as close to the meridian as possible to minimize the atmospheric effects The declination of NGC 752 is near +38$\arcdeg$ which helped to reduce those effects. There are three CCDs in the HIRES detector and the blue one has a quantum efficiency of 93\% at 3130 \AA. This is an important factor in lowering the exposure times at these short wavelengths. Our spectra covered the region of 3000$-$5800 \AA{} and have a spectral resolution of 0.027 \AA{} pix$^{-1}$ near the Be lines and a resolution element of 2 pix.

Table \ref{tab:t1} shows the properties of the stars we have observed and the details of our Keck/HIRES Be observations. The stars are designated by both their Heinemann numbers (H) \citep{heinemann1926} and by their Platais numbers (Pla) \citep{platais1991} in all the tables. We have used the value of E$(B-V)$ = 0.035 from \citet{daniel1994}. 

\begin{figure}
\centering
\includegraphics[width=\linewidth]{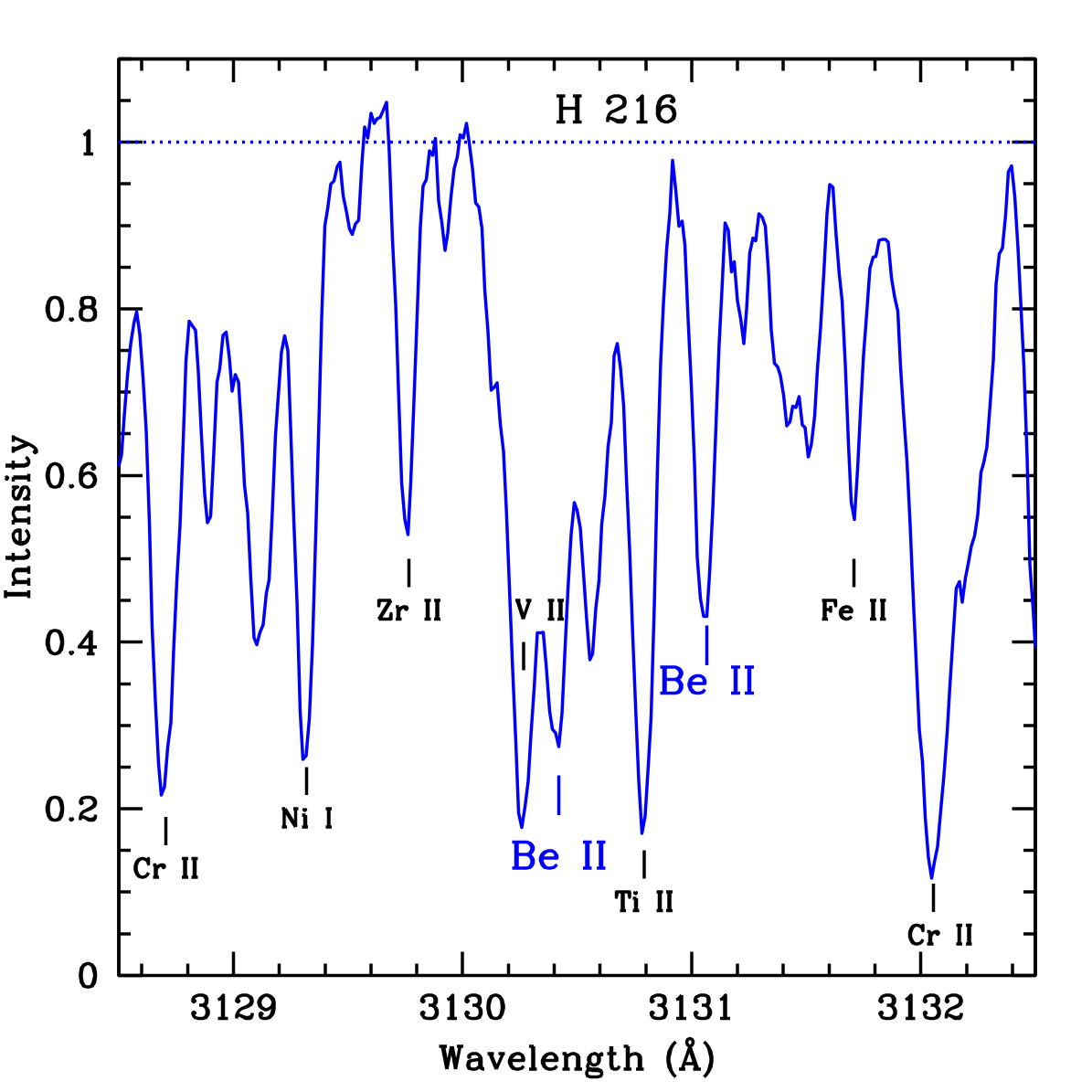}
\caption{A four \AA{} spectral region near the Be II resonance lines in one of our stars, H 216. The positions and identifications are shown for several lines in this very blended region of the spectrum.}
\label{fig:f2}
\end{figure}

An example of the Be II region in one of our stars, H 216, is shown in Figure \ref{fig:f2}. We indicate the positions of the two resonance lines of Be II as well as the identifications of several other features in that 4 \AA{} region. With a temperature of 6182 K this star is representative of several stars in our Be selection. It has a v sin i value of 4.2 km s$^{-1}$.

\begin{deluxetable*}{rrclllccc}
\tablecaption{Log of the Keck/HIRES Be Observations in NGC 752 Dwarfs \label{tab:t1}}
\renewcommand{\tabcolsep}{4mm}
\startdata
\vspace{-0.05cm}
& & & & & & & & \\
\vspace{-0.05cm}
H & Pla & $V$ & $B-V$ & ($B-V$)$_0$ & UT Date & Exp (min) & Total (min) & S/N \\
\noalign{\smallskip}
\hline
\noalign{\smallskip}
10  & 305  & 10.152 & 0.411 & 0.376 & 2019 Dec 03 & 4$\times$30 & 120 & 53 \\
80  & 520  & 12.85  & 0.57  & 0.535 & 2014 Dec 27 & 2$\times$30 & \phn60 & -- \\
-- & -- & -- & -- & -- & 2017 Nov 10 & 4$\times$30 & 120 & 49 \\
120 & 645 & 12.108 & 0.585 & 0.550 & 2019 Dec 03 & 55   & \phn55 & -- \\
-- & -- & -- & -- & --  & 2019 Dec 04 & 2$\times$45 & \phn90 & 66 \\
144 & 699 & 13.001 & 0.627 & 0.592 & 2017 Nov 11 & 5$\times$30 & 150 & 55 \\
146 & 701 & 13.06  & 0.69  & 0.655 & 2019 Dec 04 & 3$\times$45 & 135 & 34 \\
185 & 790 & 12.267 & 0.529 & 0.494 & 2017 Nov 11 & 5$\times$30 & 150 & 41 \\
197 & 824 & 11.629 & 0.440 & 0.405 & 2019 Dec 03 & 30, 42 & \phn72 & 22 \\ 
207 & 859 & 13.20  & 0.67  & 0.635 & 2019 Dec 04 & 45, 49 & \phn94 & 39 \\
216 & 889 & 12.802 & 0.551 & 0.516 & 2017 Nov 11 & 4$\times$30 & 120 & 45 \\
259 & 1000 & 11.410 & 0.419 & 0.384 & 2014 Jan 16 & 3$\times$25 & \phn75 & 59 \\
264 & 1011 & 12.128 & 0.45  & 0.415 & 2014 Dec 27 & 4$\times$25 & 100 & 51 \\
293 & 1083 & 11.92  & 0.484 & 0.449 & 2017 Nov 10 & 4$\times$30 & 120 & 56 \\
302 & 1123 & 11.507 & 0.417 & 0.382 & 2014 Jan 16 & 4$\times$25 & 100 & 29 \\
-- & 1012 & 12.417 &0.539 & 0.504 & 2017 Nov 11 & 4$\times$30 & 120 & 26 \\
\vspace{-0.3cm}
\enddata
\end{deluxetable*}

\subsection{Lithium}

As part of a larger abundance study of main-sequence stars in NGC 752, \citep{lum2019} obtained spectra in the Li I region of 10 stars. The details of the observations for Li in those stars is given in Table \ref{tab:t1}. For the Li spectra our exposure times were short and resulted in higher S/N ratios than achieved for the Be observations. These spectra cover approximately 5700$-$8120 \AA{} with some interorder gaps. The spectral resolution is 0.046 \AA{} pix$^{-1}$ near the Li I doublet, with a 2 pix resolution element.

\begin{figure}
\centering
\includegraphics[width=\linewidth]{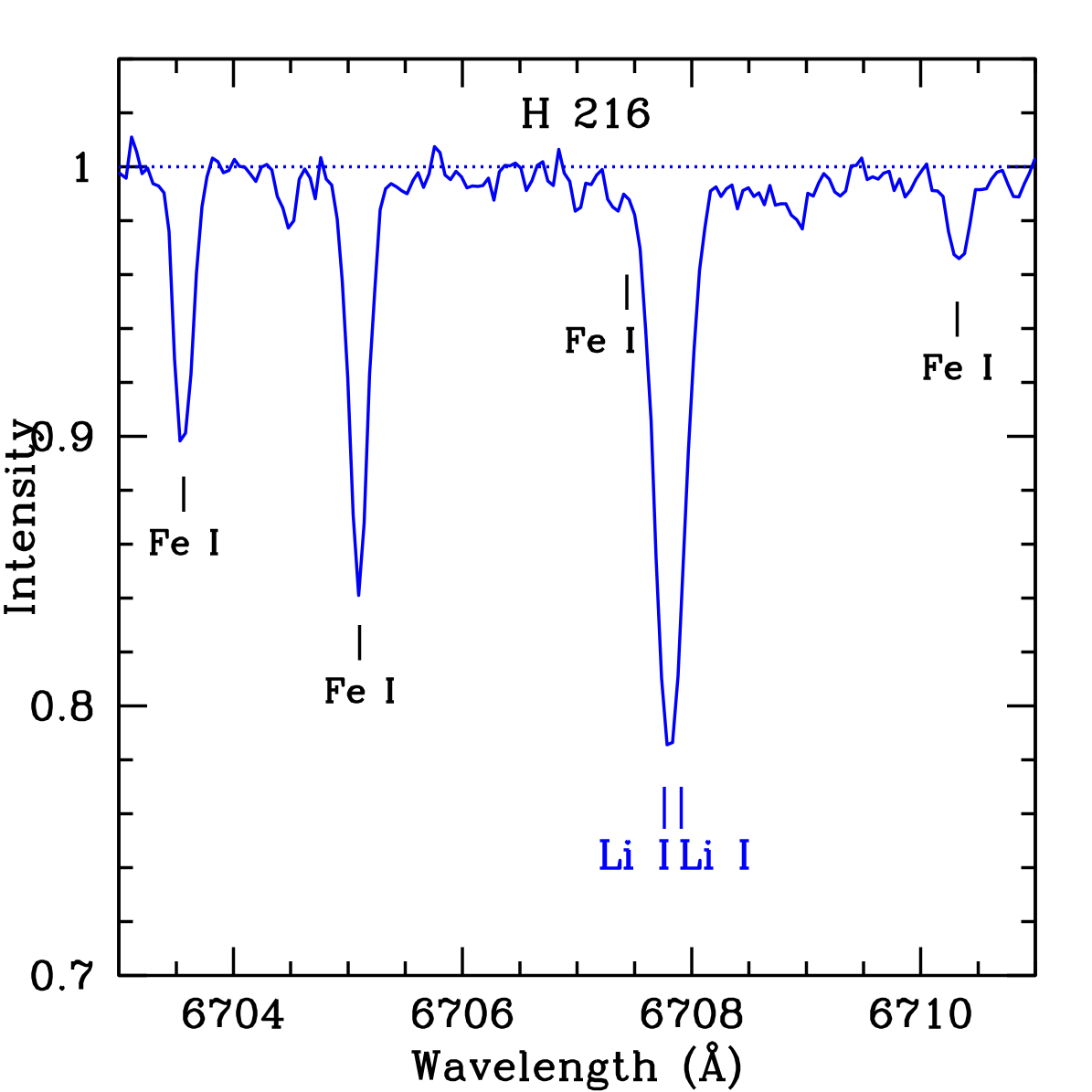}
\caption{An eight \AA{} spectral region near the Li doublet in one of our stars, H 216. The positions and identifications are shown for the Li doublet and several Fe I lines.}
\label{fig:f3}
\end{figure}

We show a 10 \AA{} spectral region of the Li I resonance doublet in Figure \ref{fig:f3} with the nearly Fe I lines indicated; this is the same star as in Figure \ref{fig:f2}. This spectral region is quite uncrowded in comparison to the Be II region. There is, however, a line of Fe I line that is found on the shortward side of the Li line and becomes stronger with decreasing temperature.

\begin{deluxetable*}{rrclllcc}
\tablecaption{Log of the Keck/HIRES Li Observations in NGC 752 Dwarfs \label{tab:t2}}
\renewcommand{\tabcolsep}{4mm}
\startdata
\vspace{-0.05cm}
& & & & & & & \\
\vspace{-0.05cm}
H & Pla & $V$ & $B-V$ & ($B-V$)$_0$ & UT Date & Exp (min) & S/N \\
\noalign{\smallskip}
\hline
\noalign{\smallskip}
80  & 520 & 12.85  &  0.57  & 0.535 & 2003 Nov 2  & 30 & 130 \\
144 & 699 & 13.00  &  0.627 & 0.592 & 2003 Nov 2  & 35 & 137 \\
146 & 701 & 12.06  &  0.69  & 0.655 & 2003 Nov 2  & 15 & \phn84 \\
184 & 791 & 12.684 &  0.543 & 0.508 & 2003 Nov 2  & 25 & 129 \\  
185 & 790 & 12.267 &  0.529 & 0.494 & 2003 Nov 2  & 20 & 139 \\
211 & 864 & 12.885 &  0.578 & 0.543 & 2003 Nov 2  & 30 & 131 \\
216 & 889 & 12.802 &  0.551 & 0.516 & 2003 Nov 2  & 30 & 136 \\
229 & 921 & 12.644 &  0.553 & 0.518 & 2003 Nov 2  & 28 & 140 \\
244 & 964 & 12.912 &  0.582 & 0.547 & 2003 Nov 2  & 25 & 117 \\
-- & 1012& 12.417 &  0.539 & 0.504 & 2003 Nov 2  & 20 & 130 \\
\vspace{-0.3cm}
\enddata
\end{deluxetable*}

We have observed Li in eight stars in common with \citet{sestito2004}. Our results for A(Li) are in excellent agreement. They range from $-$0.09 to +0.01 with a mean difference of 0.015.

\section{Abundances} 

\subsection{Stellar Parameters} \label{sec:atmos}

To determine stellar abundances it is necessary to characterize the stellar atmospheres with effective temperature, gravity, and microturbulent velocity. For NGC 752 we evaluated the parameters in much the same way as done by \citet{maderak2013}. The (B-V) photometry of \citet{daniel1994} was used along with their value of $E(B-V)$ = $-$0.035. We used the value of [Fe/H] = $-$0.01 from \citet{lum2019}. The derived temperature does depend on [Fe/H] and \citet{maderak2013} used $-$0.06, so our derived temperatures are slightly different from theirs. In particular, for the nine stars in common our temperatures are higher by 6-8 K. We used the values of log g from \citet{maderak2013} for the nine stars in common; for others we used the linear relationship between temperature and log g to find log g. We also made use of the microturbulent velocity relationship found empirically by \citet{edvardsson1993}; all of our stars were within the boundaries of their parameters for the equation for $\xi$.

Table \ref{tab:t3} gives the stellar parameters for both our Li and Be stars. For 15 of the 18 stars observed it also includes values for v sin i as listed by \citet{mermilliod2009}. We used the program \texttt{MOOG} \citep{sneden1973,sobeck2011} in {\it synthe} mode with interpolated \citet{kurucz1993} model atmospheres to find both Li and Be abundances.

We have made comparisons for temperature with other recent published work. The agreement is satisfactory. \citet{lum2019}: 10 stars +15$\pm$45; \citet{castro2016}: 5 stars +33$\pm$71; \citet{maderak2013}: 9 stars +6$\pm$1; \citet{sestito2004}: 9 stars $-$38$\pm$14. We also made log g comparisons, but only those first two references give log g values. There is excellent agreement; \citet{lum2019}: 10 stars +0.02$\pm$0.04 and \citet{castro2016}: 5 stars 0.02$\pm$0.02.

\begin{figure}
\centering
\includegraphics[width=\linewidth]{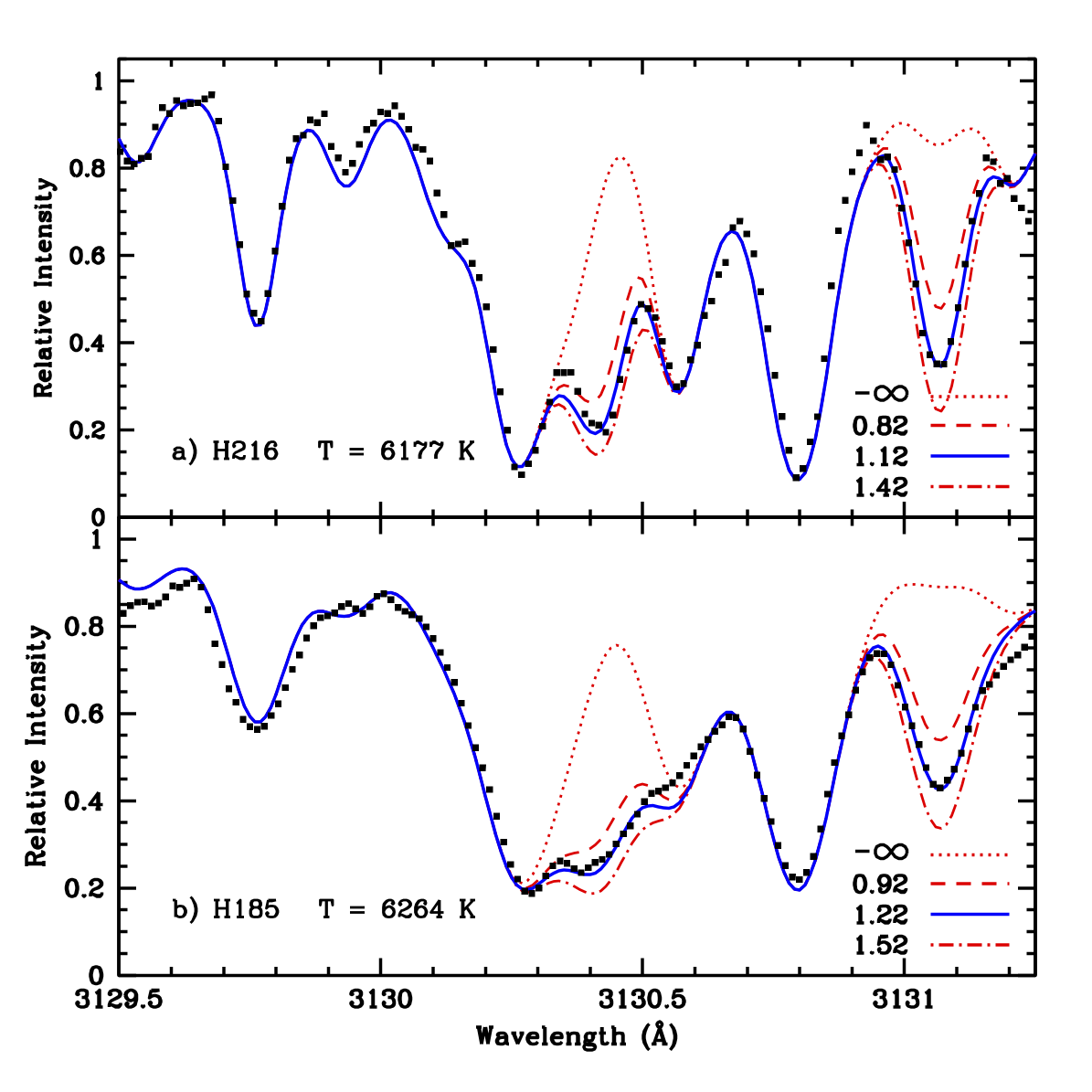}
\caption{Spectrum synthesis fits for two of our stars in the Be region. The black dots are the observations; the best fit is shown by the solid blue line. A factor of two more Be is shown by the red dotted-dashed line and a factor of two less by the red dashed line. The synthesis with no Be at all is
represented by the dotted red line.}
\label{fig:f4}
\end{figure}

\begin{figure}
\centering
\includegraphics[width=\linewidth]{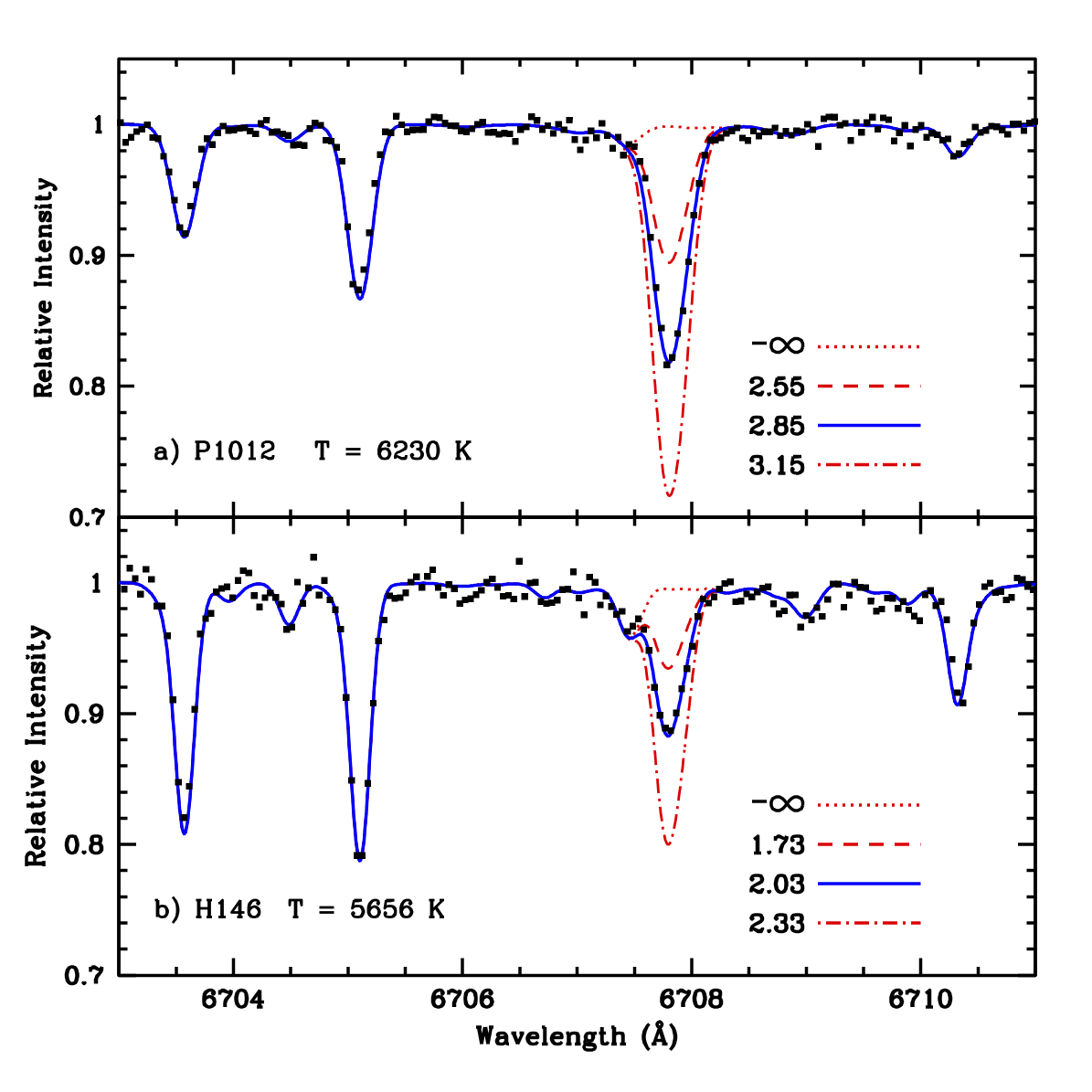}
\caption{Spectrum synthesis fits for two of our stars in the Li region. The black dots are the observations; The best fit is shown by the solid blue line. A factor of two more Li is shown by the red dotted-dashed line and a factor of two less by the red dashed line. The synthesis with no Li at all is represented by the dotted red line.}
\label{fig:f5}
\end{figure}

\begin{deluxetable}{rrcccc}
\tablecaption{Stellar Parameters for the Stars Observed in NGC 752 \label{tab:t3}}
\startdata
\vspace{-0.05cm}
& & & & & \\
\vspace{-0.05cm}
H & Pla & T$_{\rm eff}$ (K) & log $g$ & $\xi$ (km s$^{-1}$) & $v$ sin $i$ (km s$^{-1}$) \\
\noalign{\smallskip}
\hline
\noalign{\smallskip}
10   & 305   & 6759 & 4.23 & 2.21 & 17.5 \\
80   & 520   & 6102 & 4.43 & 1.43 & \phn0.4 \\
120  & 645   & 6246 & 4.39 & 1.59 & \phn9.6 \\
144  & 699   & 5893 & 4.48 & 1.18 & \phn2.9 \\
146  & 701   & 5656 & 4.53 & 0.94 & -- \\
184  & 791   & 6208 & 4.40 & 1.55 & \phn7.3 \\
185  & 790   & 6264 & 4.38 & 1.62 & \phn9.2 \\
197  & 824   & 6633 & 4.27 & 2.06 &  28.4 \\
207  & 859   & 5735 & 4.52 & 1.01 &  -- \\
211  & 864   & 6071 & 4.44 & 1.39 &  \phn3.4 \\
216  & 889   & 6177 & 4.41 & 1.51 &  \phn4.2 \\
229  & 921   & 6199 & 4.41 & 1.51 &  \phn2.6 \\
244  & 964   & 6063 & 4.44 & 1.39 &  \phn3.6 \\
259  & 1000  & 6724 & 4.24 & 1.05 &  15.3 \\
264  & 1011  & 6635 & 4.27 & 2.06 &  32.7 \\
293  & 1083  & 6452 & 4.33 & 1.83 &  19.7 \\
302  & 1123  & 6703 & 4.25 & 2.14 &  34.0 \\
-- & 1012 & 6230 & 4.39 & 1.55 &  \phn6.5 \\
\vspace{-0.3cm}
\enddata
\end{deluxetable}

\subsection{Beryllium}

Our synthesis covered the spectral region from 3129.5$-$3132.5 \AA{} with a line list of over 300 spectral features and includes atomic lines as well as molecular features due to CH, NH, OH, CN and CO. As can be seen in Figure \ref{fig:f2} the longward Be II line at 3131 \AA{} is relatively free of blends so we relied primarily on that line to determine the amount of Be.

Examples of the Be spectrum synthesis fits can be seen in Figure \ref{fig:f4} for two stars. The star H 216 has v sin i = 4.2 km s$^{-1}$ while H 185 in the lower panel shows rotation of more than double that at 9.2. The figure shows that for the more rapidly-rotating stars, it is more difficult to be confident of the fit for the shortward, more-blended Be II line. The results for A(Be) are given in Table \ref{tab:t4}. A(Be) = log N(Be)/N(H) + 12.00.

Unfortunately, the stars in the Li dip are the hotter ones and thus they are more likely to be rotating rapidly. There is a dichotomy in stellar rotation rates that was first shown by \citet{kraft1967} such that stars hotter than $\sim$6200 K are rotating rapidly. That larger rotation rate results in severe blending in this spectral region. Five of the stars we observed for Be are in the Li-dip. Three of the Li-dip stars -- H197, T=6633 K; H264, T=6635 K); H302 T=6703 K -- have low upper limits on A(Li) (= log N(Li)/N(H) + 12.00). For these stars we could only determine upper limits on the Be abundance also. There are two stars in the Li dip that have measurable but depleted Li. One star, H293 (T=6452 K), shows measurable, but depleted Be with A(Be) = 0.82. The other, H259 (T=6724 K), is on the hot side of the Li dip and may show mild Be depletion with A(Be) = 1.12. For one of the hotter stars, H10, we were unable to synthesize a spectrum that matched the observed in spite of a respectable S/N ratio. This star is brighter by a magnitude
than the other stars in our sample. It is classed as a spectroscopic binary in SIMBAD which would account for its brightness and complicated spectrum.

\subsection{Lithium}

In the Li region we synthesized 8 \AA{} from 6703.3 to 6711.0. Our line list had 95 lines which included some 40 CN features. Figure \ref{fig:f5} shows the synthesis for two stars. The Li abundance is sensitive to temperature and cooler stars show stronger Li lines for the same Li abundance. Here our coolest star, H 146, has a weak Li lines due to its Li depletion compared to the hotter star, Pla 1012. The blending feature at 6707.432 is due to Fe I, and can be seen clearly in the spectrum of H 146 and in the synthesis fit; this is clear in the fit containing no Li (red dotted line) which does have Fe. Our A(Li) results are also given in Table \ref{tab:t4}.

\subsection{Error Estimates}

It is possible to use \texttt{MOOG} to find abundance errors resulting from stellar parameter uncertainties. We have used models on the Kurucz-model grid with a range in temperature, log g and [Fe/H]. We used \texttt{MOOG} in the "blends" mode with 13 lines blending with the Be II lines at 3131 \AA. A summary of those results follows. A change in temperature of $\pm$80 K gives a change in A(Be) of $\pm$0.01, except there is no change in the range from 6000 to 6250 K. The largest error stems from log g where an uncertainty of +0.1 leads to an uncertainty in A(Be) from 0.036 to 0.052, increasing with decreasing temperature from 6500 to 5750 K. As [Fe/H] increases by +0.05, A(Be) increases by $\sim$+0.02. For each star for which we determine a Be abundance we have assessed errors due to each of the parameters and found the total error by summing them in quadrature. Those range from $\pm$0.042 to $\pm$0.057 with the coolest stars have the largest errors.

Another uncertainty in the determination of Be abundances actually comes from
the "goodness-of-fit" in the line profile matching (sigma calculated by \texttt{MOOG}) and from the relative reliance of the two Be II lins. The line at 3131 \AA{} has substantially fewer blending features as can be seen in Figures \ref{fig:f2} and \ref{fig:f4}. For three of the stars the rotational broadening prevented a determination of the Be abundance and only upper limits could be found: H 197, H 264, H 302. For nine others the fit for the 3131 line was excellent while the blend at the 3130 line gave a slightly lower Be abundance; we relied on the 3131 line. The fit of a continuum in this spectral region produces another uncertainty that is difficult to evaluate; experience suggests this can be done to $\pm$5\% although this varies depending on the signal-to-noise of the exposure and the rotational velocity of the star. An error estimate is presented in Table \ref{tab:t4}.

We have determined the Li abundances by spectrum synthesis also using the ``synthe'' routine in \texttt{MOOG}. The error in the Li abundance is almost entirely from the error in the effective temperature. A temperature difference of $\pm$75 K translates to $\pm$0.06 in A(Li). We synthesized a region of 8 \AA{} which includes two Fe I lines with differing excitation potentials; the fits are excellent for both of those lines and that provides another check on the temperature. The Fe I line at 6707.432 blends with the Li feature at cooler temperatures but is well-accounted for in the spectrum synthesis analysis and is not a factor in the error evaluation. The Li region spectra and the syntheses can be seen in Figure \ref{fig:f3} and \ref{fig:f5}.

\section{Discussion}

\subsection{Lithium and Beryllium in NGC 752}

We have determined Be abundances in 13 stars and Li abundances in 10 stars. Abundances of Li have been determined for an additional 29 stars in previous research. We have noted above that there are some temperature differences with those previous studies of in NGC 752. In order to make useful
comparisons we put those investigations of Li abundances on the temperature scale used here and by \citet{maderak2013} and recalculated the values of the Li abundances. Our values and these revised results for T$_{\rm eff}$ and A(Li) abundance are given in Table \ref{tab:t5} for 39 main-sequence stars in NGC 752 along with the original references.

\begin{deluxetable}{rrcrrr}
\tablecaption{Li and Be Abundance Results for NGC 752 \label{tab:t4}}
\startdata
\vspace{-0.05cm}
& & & & & \\
\vspace{-0.05cm}
H & Pla & T$_{\rm eff}$ (K) & A(Be) & $\sigma$(Be) & A(Li) \\
\noalign{\smallskip}
\hline
\noalign{\smallskip}
10   & 305   & 6759 & --        & --	   & -- 	\\
80   & 520   & 6102 & 1.29       & 0.073   & 2.72 	\\
120  & 645   & 6246 & 1.15       & 0.068   & -- 	\\
144  & 699   & 5893 & 1.27       & 0.078   & 2.40    	\\
146  & 701   & 5656 & 1.20       & 0.076   & 2.03    	\\
184  & 791   & 6208 & --    & -- & 2.58   	\\
185  & 790   & 6264 & 1.20       & 0.069   & 2.83 	\\
197  & 824   & 6633 & $\leq$0.1  & -- & -- 	\\
207  & 859   & 5735 & 1.10       & 0.076   & -- 	\\
211  & 864   & 6071 & --    & -- & 2.73 	\\
216  & 889   & 6177 & 1.15       & 0.071   & 2.87 	\\
229  & 921   & 6199 & --    & -- & 2.75 	\\
244  & 964   & 6063 & --    & -- & 2.68 	\\
259  & 1000  & 6724 & 1.12       & 0.065  & -- 	\\
264  & 1011  & 6635 & $\leq$0.40 & -- & -- 	\\
293  & 1083  & 6452 & 0.82       & 0.065  & -- 	\\
302  & 1123  & 6703 & $\leq$0.12 & -- & --    \\
-- & 1012 & 6230 & 1.20       & 0.068   & 2.85 	\\
\vspace{-0.3cm}
\enddata
\end{deluxetable}

\begin{deluxetable}{rrcrr}
\tablecaption{Lithium Results on a Common Scale in NGC 752 Dwarfs \label{tab:t5}}
\renewcommand{\tabcolsep}{4mm}
\startdata
\vspace{-0.05cm}
& & & & \\
\vspace{-0.05cm}
H & Pla & $T_{\mathrm{eff}}$ & A(Li) & Reference$^{*}$ \\
\noalign{\smallskip}
\hline
\noalign{\smallskip}
10 & 305 & 6759 & 2.86 & HP86, B95	\\
80 & 520 & 6102 & 2.72 & B22, S04	\\
87 & 552 & 6066 & 2.64 & S04	\\
88 & 555 & 6460 & $<$2.01 & PH88, B95	\\
94 & 575 & 5423 & $<$1.0 & HP86, B95, S04	\\
120 & 648 & 6246 & 2.29 & S04	\\
123 & 654 & 6824 & $<$2.73 & HP86, B95	\\
139 & 689 & 6572 & $<$2.10 & PH88, B95	\\
144 & 699 & 5893 & 2.40 & B22	\\
146 & 701 & 5656 & 2.03 & B22, HP86, S04	\\
183 & 786 & 5524 & 1.49 & S04	\\
184 & 791 & 6208 & 2.58 & B22	\\
185 & 790 & 6264 & 2.83 & B22, HP86, S04	\\
189 & 799 & 6708 & 3.20 & PH88, B95	\\
197 & 824 & 6633 & $<$2.09 & PH88, B95	\\
207 & 859 & 5735 & 0.98 & HP86, B95, S04	\\
211 & 864 & 6071 & 2.73 & B22, S04	\\
216 & 889 & 6177 & 2.87 & B22, S04	\\
222 & 901 & 6859 & 3.08 & HP86, B95	\\
229 & 921 & 6199 & 2.75 & B22, HP86, S04	\\
235 & 950 & 6468 & $<$1.12 & PH88, C16	\\
237 & 953 & 5994 & 2.39 & HP86, B95, S04	\\
244 & 964 & 6063 & 2.68 & B22, S04	\\
252 & 983 & 5956 & 2.38 & S04	\\
254 & 988 & 6925 & 3.17 & HP86, B95	\\
256 & 993 & 5601 & 1.57 & S04	\\
258 & 998 & 6302 & 2.27 & HP86	\\
259 & 1000 & 6724 & 2.49 & HP86, B95	\\
264 & 1011 & 6635 & $<$1.87 & HP86	\\
265 & 1017 & 5771 & 1.59 & HP86, B95, S04	\\
266 & 1023 & 6703 & 2.93 & HP86, B95	\\
293 & 1083 & 6452 & 2.22 & HP86, B95	\\
298 & 1107 & 5777 & 1,87 & S04	\\
302 & 1123 & 6703 & $<$1.82 & HP86, B95	\\
304 & 1129 & 6464 & $<$1.16 & PH88, B95	\\
-- & 475 & 6903 & 2.48 & S04	\\
--	& 1012 & 6230 & 2.85 & B22, S04	\\
--	& 1284 & 5801 & 2.20 & S04	\\
--	& 1365 & 5593 & 2.06 & S04	\\
\vspace{-0.3cm}
\enddata
\tablecomments{\\
$^{*}$ B22 = this work; B95 = \citealt{balachandran1995}; S04 = \citealt{sestito2004}; HP86 = \citealt{hobbs1986}; PH88 = \citealt{pilachowski1988}; C16 = \citealt{castro2016}}
\end{deluxetable}

\begin{figure}
\centering
\includegraphics[width=\linewidth]{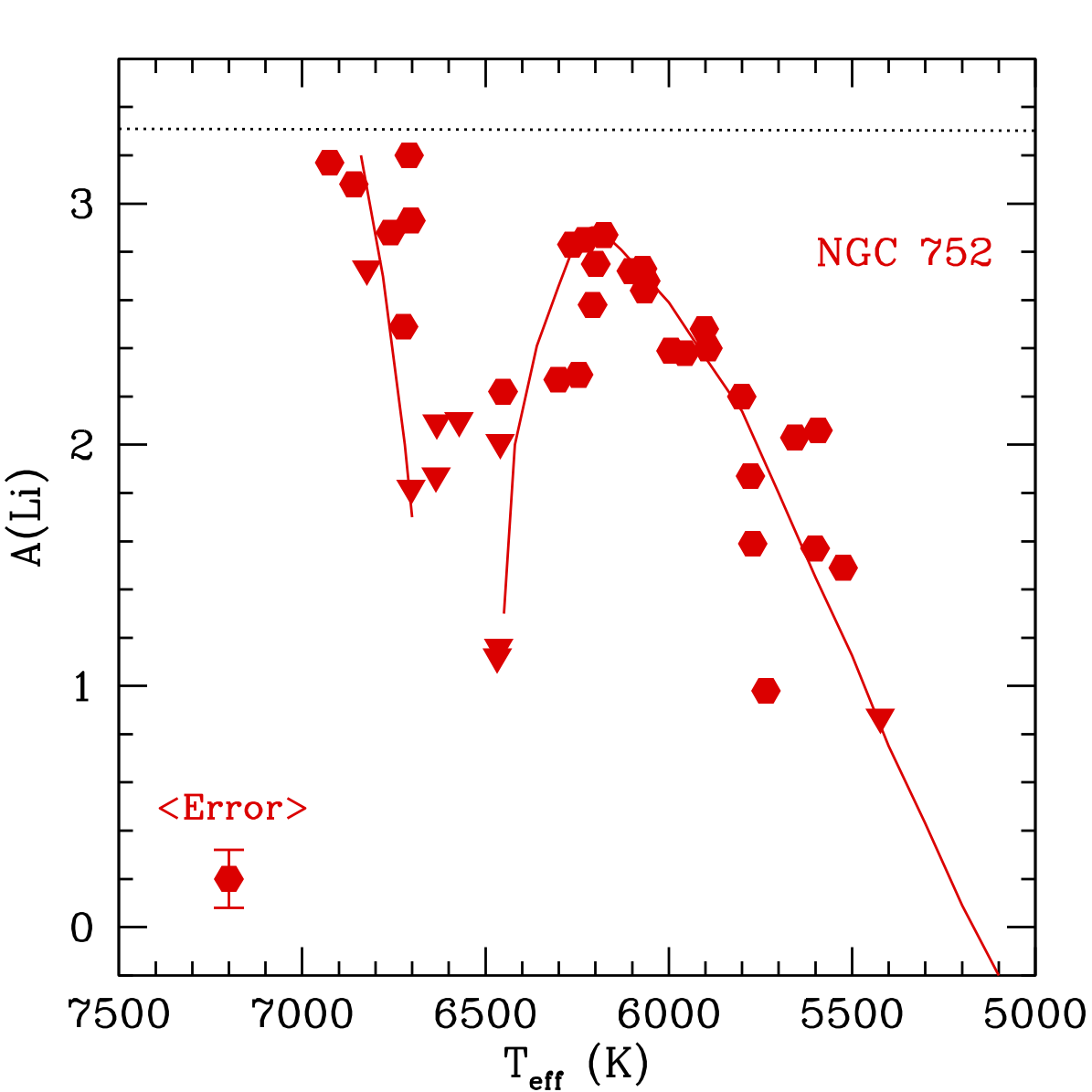}
\caption{Lithium abundances in NGC 752 for 39 stars on a common temperature scale. The red line is a fit through the data points and shows the Li dip near 6500 K which is typical of several open clusters. A mean error bar on A(Li) is shown in the lower left corner.}
\label{fig:f6}
\end{figure}

Figure \ref{fig:f6} shows those results for Li together with an approximate fit through to points. This fit is modeled after the fit through the Li-temperature relation for the Hyades showing an unmeasurably deep Li dip near 6500 K, a peak in Li near 6200 K followed by a steep decline to vanishingly small amounts of Li near 5200 K.

\begin{figure}
\centering
\includegraphics[width=\linewidth]{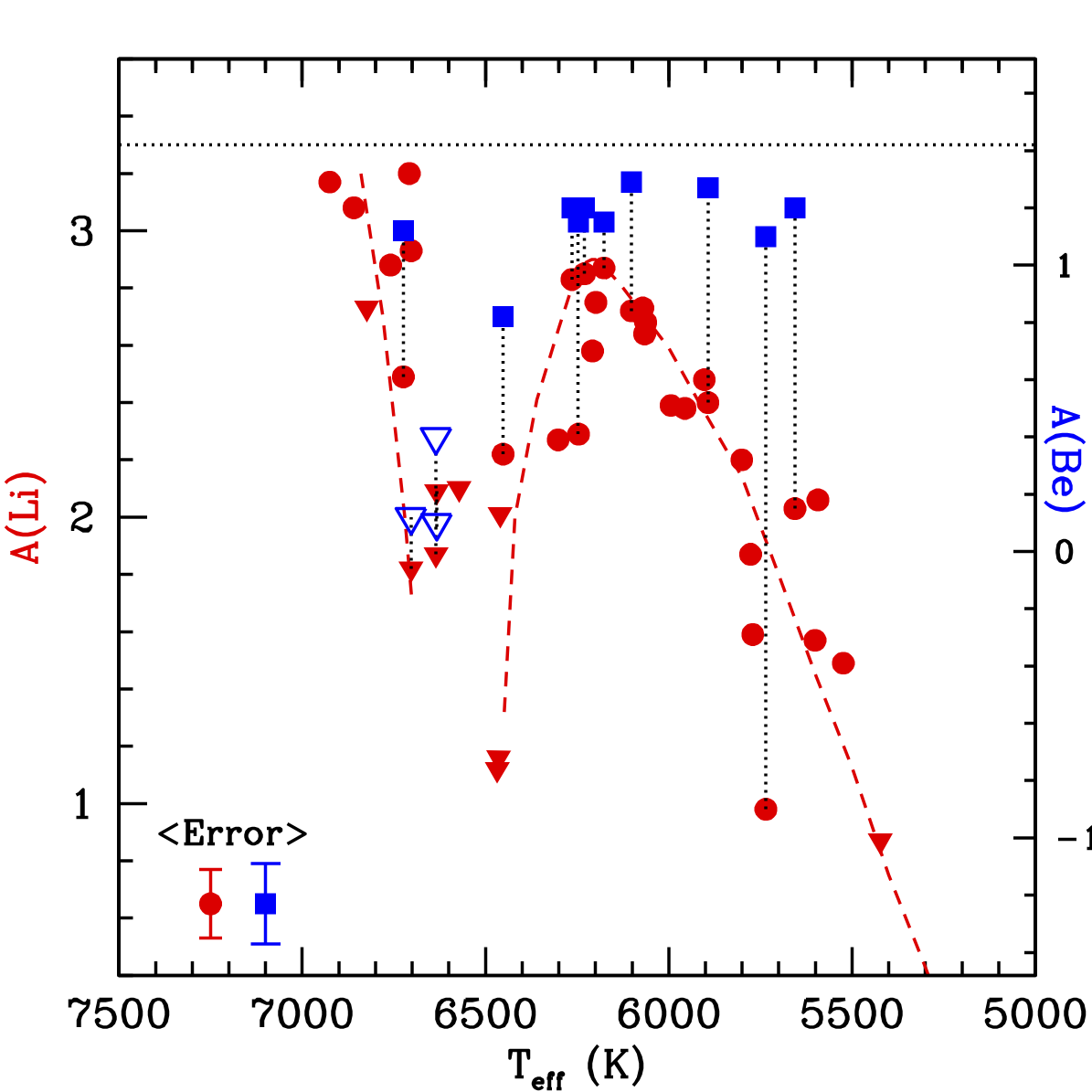}
\caption{Abundances of Li and Be in NGC 752 shown of their respective vertical axes, on the same scale and normalized to their respective meteoritic content. The Li values are filled red circles with upper limits on Li as filled red triangles. The Be points are filled blue squares with upper limits as open blue triangles. Results for the same stars are connected by dotted vertical lines. The Li fit is shown by the dashed red line. Typical error bars for Li and Be are shown in the lower left corner.}
\label{fig:f7}
\end{figure}

The results for Be can be added to our diagram in Figure \ref{fig:f6} with the outcome for Li on the same vertical scaling but referenced to their individual meteoritic abundances from \citet{anders1989} (see the discussion about Be in the solar system by \citealt{lodders2003} who promotes the need to clarify the solar Be and the need for more Be analyses in Ci chondrites). This is shown in Figure \ref{fig:f7}. Our Be values for the stars in NGC 752 are connected by vertical dotted lines to their Li values for easier reference.

The four stars in the Li-dip region are clearly depleted in Be with three showing only upper limits for both Li and Be and the fourth (H 293) showing A(Li) = 2.22 and measurable, but depleted A(Be) at 0.82. They show strong evidence for a Be-dip. A fifth star, H259, at T = 6724 K, with measurable Li depletion shows marginal Be depletion with A(Be) = 1.12.

For the solar-like stars with effective temperatures below $\sim$6400 K there is little Be depletion. The data in Table 4 give a mean value for A(Be) in those eight stars of 1.20 $\pm$0.06. This is comparable to the solar Be value of 1.15 $\pm$0.20 from the careful study of solar Be by \citet{chmielewski1975} and of the 1.15 value found by \citet{boesgaard2002} from a daytime sky Be spectrum with the same telescope and instrument as our data for NGC 752. \citet{king1997} found a probable Be abundance in the Sun in the range A(Be) = 0.91$-$1.26 corresponding to a probable Be depletion of 0.16 to 0.50. In a recent paper \citet{asplund2021} use a 3-D hydrodynamic model of the Sun to determine the chemical make-up of the Sun. They give the solar Be as 1.38 $\pm$0.09 but suggest further analysis with a careful reassessment of the blends.

Figure \ref{fig:f7} also shows how small the Be depletions in NGC 752 are relative to their Li depletions in the three coolest stars. Those Li depletions are substantial in the cooler stars.

\subsection{Comparison with the Hyades}

In order to make comparisons with the younger Hyades cluster, we need to insure that the stellar parameter scales are alike. For NGC 752 the temperature scale is consistent with what was used for the Hyades in the Li, Be, B research by \citet{boesgaard2016}. They followed the study of \citet{debruijne2001}, who used the results from multiple measurements of $(B-V)$ from {\it Hipparcos} for 92 ``high-fidelity'' Hyades members. They used the calibrations of \citet{bessell1998} to convert measured (B-V) values to T$_{\rm eff}$; they correct for the Hyades metallicity of +0.14 following \citet{alonso1996}. The values for T$_{\rm eff}$ in the Hyades
are given in Table 3 in the study of \citet{boesgaard2016}.

Another large study that included Li abundance in the Hyades was done by \citet{cummings2017}. They used many (ground-based) color indices to find stellar temperatures. They converted each color index to an effective B-V index and used the calibration described in \citet{deliyannis2002} to find temperatures for their 90 Hyades stars. The temperature agreement with the 79 Hyades stars from \citet{boesgaard2016} is remarkably good; for the 73 stars in common the slope of the relationship of the temperatures is 1.0094. 

\begin{figure}
\centering
\includegraphics[width=\linewidth]{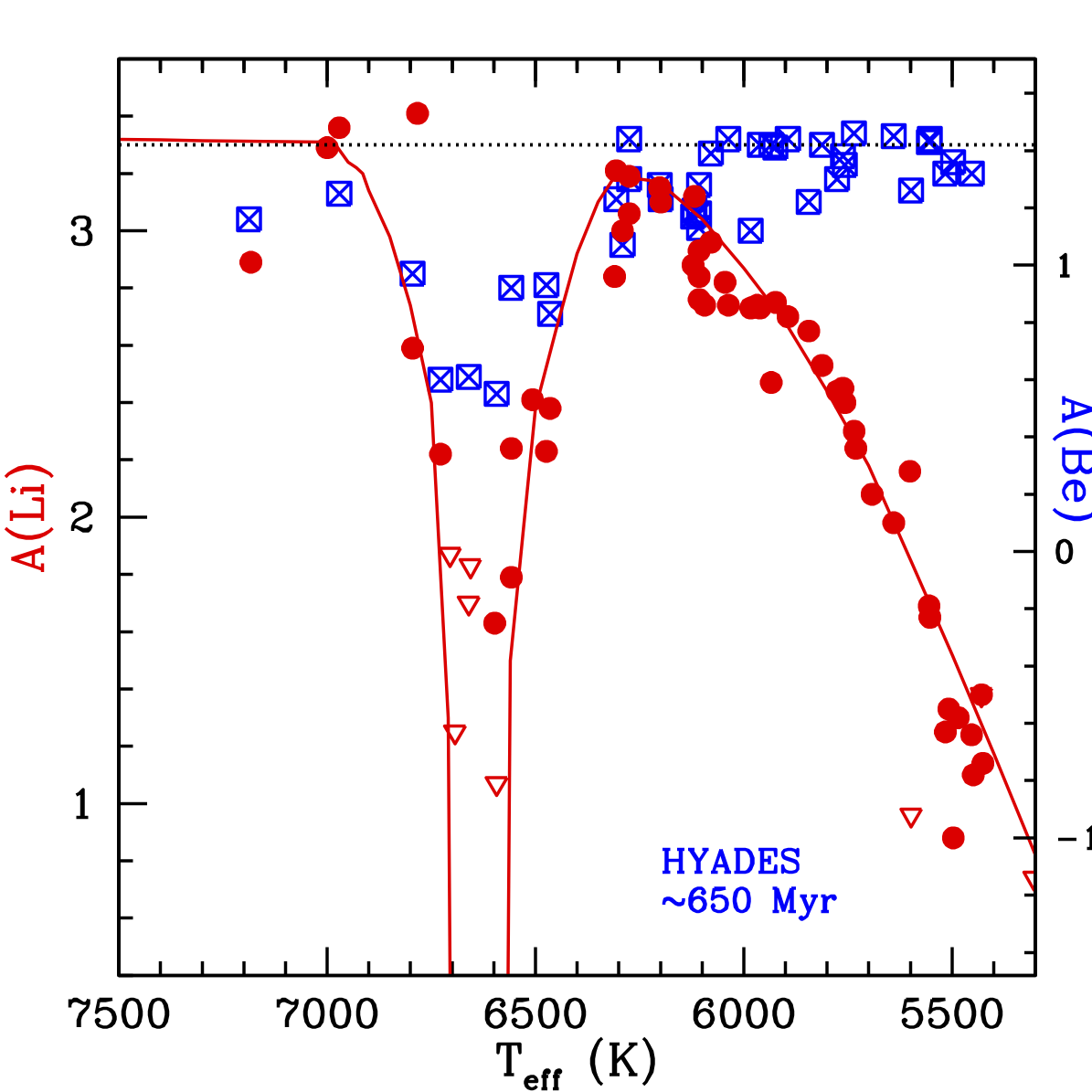}
\caption{Similar to Figure \ref{fig:f7}, but for the Hyades cluster. The Li abundances
are shown as filled red circles, with upper limits as open red triangles. The Be abundances are crossed blue squares. The solid red line is the fit to the Hyades Li data.}
\label{fig:f8}
\end{figure}

Those two Hyades Li studies have the same temperature scale. The method from \citet{cummings2017} for the Hyades temperature determination is the same one used by \citet{maderak2013} for NGC 752. Here we have used that same method for temperation determination with the exception of the value for [Fe/H] for NGC 752; this was discussed in Section \ref{sec:atmos}. The log g scales are identical. We can conclude that the scales for NGC 752 and the Hyades are as similar as possible.

The counterpart diagram to Figure \ref{fig:f7} for Li and Be in the Hyades in shown in Figure \ref{fig:f8} which is based on the Li and Be results of \citet{boesgaard2016}. The younger Hyades cluster shows a deep Li dip centered at $\sim$6650 K and a clear matching Be dip. In the cooler dwarfs there is little Be depletion; we have found little cool star depletion of Be for the older NGC 752. This is similar to the earlier Hyades findings of \citet{boesgaard2002} who reported a mean A(Be) = 1.33 $\pm$0.06 for the 10 coolest stars and to the more recent result of \citet{boesgaard2016} for the mean of the 31 stars below 6400 K of A)Be) = 1.31 $\pm$0.12. The benchmark Be abundance for the solar system, as measured in meteorites, is 1.42 $\pm$0.04 \citep{anders1989}.

The Hyades cluster is younger than the Sun and mildly metal-rich at [Fe/H] = +0.13 so one might ask if it has initial values of Li and Be that are higher than the solar system values: A(Li) = 3.31 and A(Be) = 1.42. The stars in the Hyades and its sister cluster, Praesepe, have been well studied for Li abundances: \citet{boesgaard1988}, 35 stars with none higher than A(Li) = 3.30; \citet{takeda2013}, 68 stars with the (one) highest at 3.25; \citet{boesgaard2016}, 79 stars with one at 3.41; \citet{cummings2017}, 90 stars with one (that same star) at 3.33 and in Praesepe, 110 stars with only one at 3.33. Although they did not study the Hyades or Praesepe, \citet{randich2020} found Li abundances in 18 open clusters with four having [Fe/H] between 0.20
and 0.26; those four clusters showed a mild increase in A(Li) to 3.4. Beryllium is less well studied but in the Hyades \citet{boesgaard2016} determined A(Be) in 43 stars with most of the cool stars having Be values near 1.42 (e.g., see Figure \ref{fig:f8}). We can conclude that the meteoritic abundances provide a useful comparison as initial abundances for Li and Be. Furthermore, there are virtually no F, G, K field stars with higher Li values than 3.31. \citet{ramirez2012} present Li abundances for 1381 FGK dwarfs and subgiants; only ten of those stars had A(Li) $>$3.3.

Although more than half the stars with T $<$6400 K show no apparent Be depletion, the Hyades mean value of 1.33 $\pm$0.06 is above that mean of 1.20 $\pm$0.06 found for eight stars in NGC 752. The Hyades cluster is $\sim$680 Myr \citep{gossage2018} while NGC 752 is about twice that age at $\sim$1.34 Gyr \citep{agueros2018} and the difference in the mean Be abundances in the G dwarfs in the two clusters could be due to the age difference; there has been more time for the larger depletion of Be to have occurred in NGC 752. Both values are below the original solar system value yet higher than the depleted solar content of Be of 1.15.

The comparison of Figures \ref{fig:f7} and \ref{fig:f8} shows that the Be dip in the older NGC 752 is deeper than the Be dip in Hyades. The Hyades Be depletion is measurable; Be is depleted by a factor of 6-7 from the meteoritic value, and, from the Hyades Be value in the cooler G dwarfs. In NGC 752, two Be detections and three upper limits on Be were found. The upper limits indicate Be depletions of more than an order of magnitude. In both clusters Li is severely depleted and unmeasurably low in the center of the dip.

For the Hyades in the temperature region around 6200 to 6400 K there is a Li ``peak.'' Those stars have A(Li) not far below the meteoritic abundance. However, in NGC 752 those stars are Li-depleted by about a factor of 3. The older age of NGC 752 contributes to this difference in Li.

In the cooler stars, the G dwarfs, Be seems uniformly depleted in NGC 752 and more depleted than in the Hyades stars relative to the meteoritic value. In the cooler stars, the Li content is decreasing with decreasing temperature in both Hyades and NGC 752, but there are larger Li depletions at a given temperature in the older NGC 752. Recall that the Be abundances appear similarly constant in this temperature range in both clusters with a apparently reduced level in NGC 752.

\begin{figure}
\centering
\includegraphics[width=\linewidth]{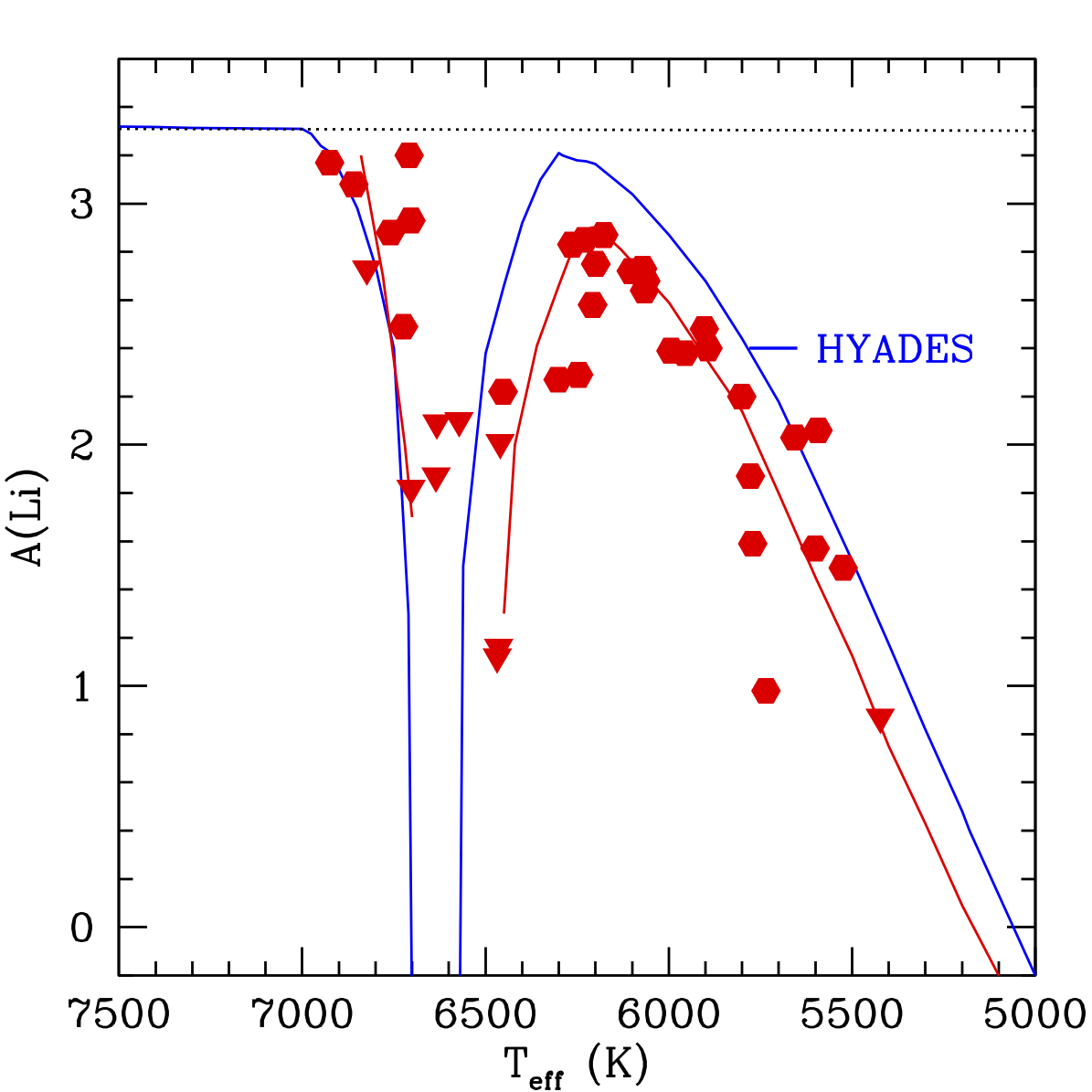}
\caption{Lithium abundances as a function of temperature. The red data points and the red line represent the values for NGC 752. The blue line represents the Li fit for the Hyades cluster.}
\label{fig:f9}
\end{figure}

In Figure \ref{fig:f9} we show the fits for both NGC 752 and the Hyades for the Li-temperature relationship along with the Li data points for NGC 752. The figure makes the Li differences between the young Hyades and the older NGC 752 stand out. Although the hot side of the Li dip seem similar in the two cluster, it is clear that the Li gap is broader in NGC 752, i.e. the cool side of the dip has advanced to cooler temperatures in the older cluster. When the breadth of the dip is measured at A(Li) = 2.5, it is 295 K in the Hyades and 480 K in NGC 752, or 185 K wider in NGC 752. We thus quantify the suspicion by \citet{hobbs1986} of ``growth of the gap with age.''

Figure \ref{fig:f9} also makes it clear that the peak Li values near 6200$-$6300 K are lower in NGC 752. For the Hyades the peak A(Li) is 3.20 compared to 2.85 for NGC 752, or a factor of $\sim$2 greater depletion in NGC 752. Although the fall-off of A(Li) toward cooler temperatures has a similar pattern in the two clusters, there is greater Li depletion at each temperature in NGC 752.

For NGC 752 we find a 1) a greater width of the Li dip toward cooler stars, 2) a deeper Be dip, 3) a lower Li peak, 4) a Be mean for the three coolest stars that may be lower than the meteoritic Be and may be lower than the Hyades mean, and 5) a greater Li depletion in the stars cooler than $\sim$6200 K at each temperature. All five of these phenomena cam be attributed, at least in part, to the older age of NGC 752. In addition to the age difference between the two clusters, there is a difference in the metallicity with the older NGC 752 having only 70\% of the metal content of the Hyades. The greater amount of the heavy elements in the Hyades stars would lead to a greater opacity which, in turn, creates a deeper convection zone down to hotter temperatures. This effect could lessen the impact of the age difference.

The most successful theory to account for the simultaneous depletions of the light elements in the Li-Be dip appears to be extra mixing below the surface convection zone caused by stellar rotation and spin-down of that rotation with time, e.g. \citet{charbonnel1994}, \citet{deliyannis1989},
\citet{somers2016}, \citet{dumont2021}. 

\begin{figure}
\centering
\includegraphics[width=\linewidth]{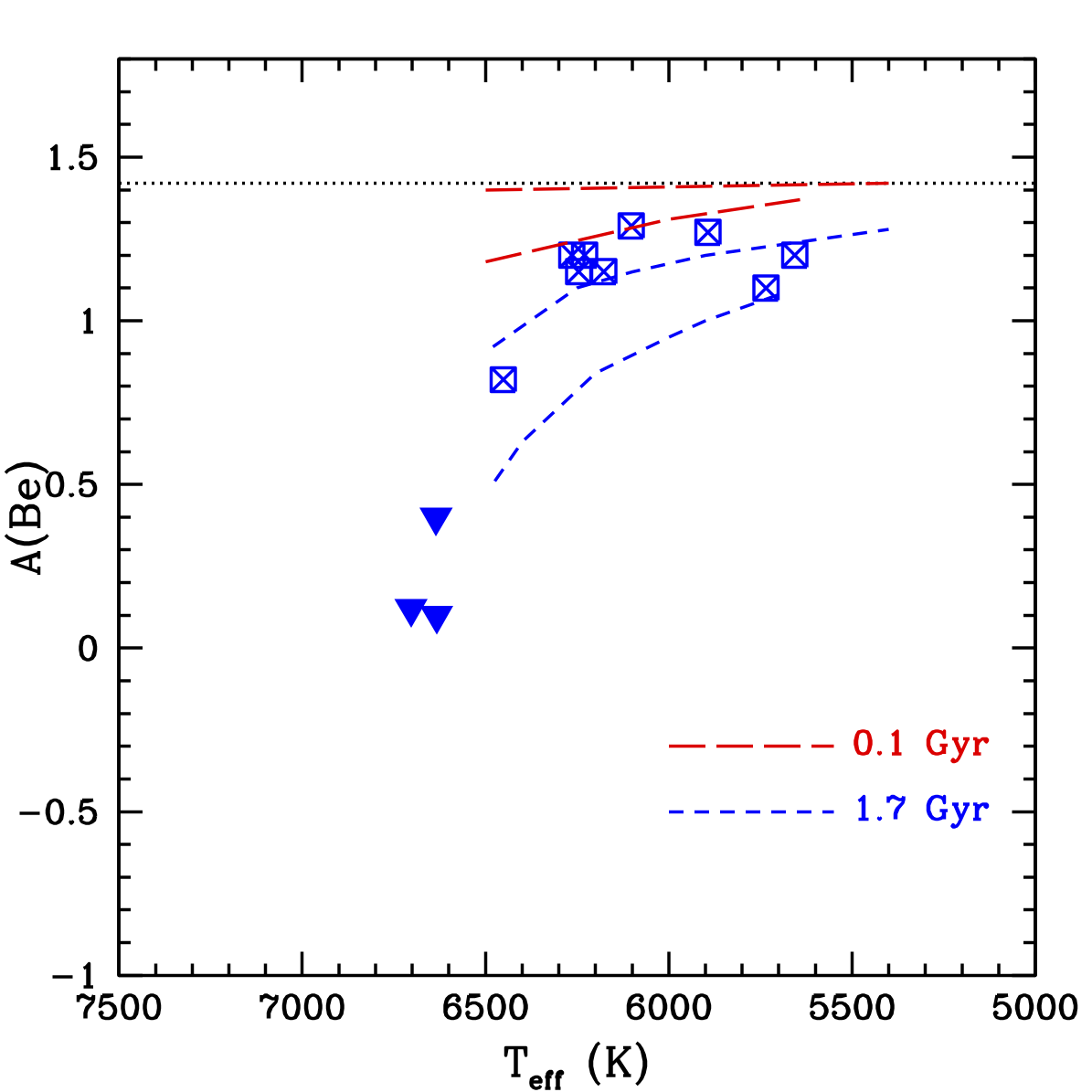}
\caption{Results for A(Be) in NGC 752 with temperature shown as blue
crossed squares and Be upper limits as filled triangles. The long red dashes correspond to theoretical calculations for 0.1 Gyr and the short blue dashes to 1.7 Gyr. In each case the upper curves are for initial rotational velocities of 10 km s$^{-1}$ and the lower ones to 30 km s$^{-1}$.}
\label{fig:f10}
\end{figure}

These models predict the dependences of Li and Be with temperature that we observe in these two clusters, and that an older cluster should be more depleted in both elements. The models take into account that stars spin down during the main sequence, which triggers a shear instability that causes mixing and thus surface depletion of Li, Be, and B. The degree of depletion depends on how much angular momentum is lost, and continues with age as stars continue to spin down. In going from G dwarfs to F dwarfs (increasing in mass), stars form with an increasing amount of initial angular momentum and lose more as they age \citep{barnes2007,reinhold2015}. This creates the F-dwarf Li and Be dips, and the Li peak (less depletion) in late-F/early-G dwarfs. However, in increasingly cooler G dwarfs the surface convection zone occupies an increasingly substantial fraction of the Li preservation region, so increasingly less rotational mixing is needed to deplete Li. Thus the Li
abundances decline steeply, in spite of the smaller initial angular momenta in these stars. However, the predicted pattern for Be is different. Since the surface convection zone is tiny in F-dwarfs compared to the (Li and) Be preservation region(s), rotational mixing does nearly all the mixing, creating a Be dip. However, in going to cooler stars, the Be preservation region remains substantially deeper than the surface convection zone, so there is no Be peak, just a flattening of the Be-temperature relation, with little Be depletion in the cooler G dwarfs.

One of the five interesting results enumerated above is that the Be abundances in the cooler stars in NGC 752 appear to be lower than those in the Hyades in stars of similar temperature; compare Figures \ref{fig:f7} and \ref{fig:f8}. In fact, this agrees with previous published results. In particular, \citet{stephens1997} show calculations of both Li and Be deletions in their Figures 11 and 12. The depletions are due to rotationally-induced mixing. They were calculated for initial equatorial rotation speeds of 30 and 10 km s$^{-1}$ and for ages of 0.1, 1.7 and 4.0 Gyr.

\begin{figure}
\centering
\includegraphics[width=\linewidth]{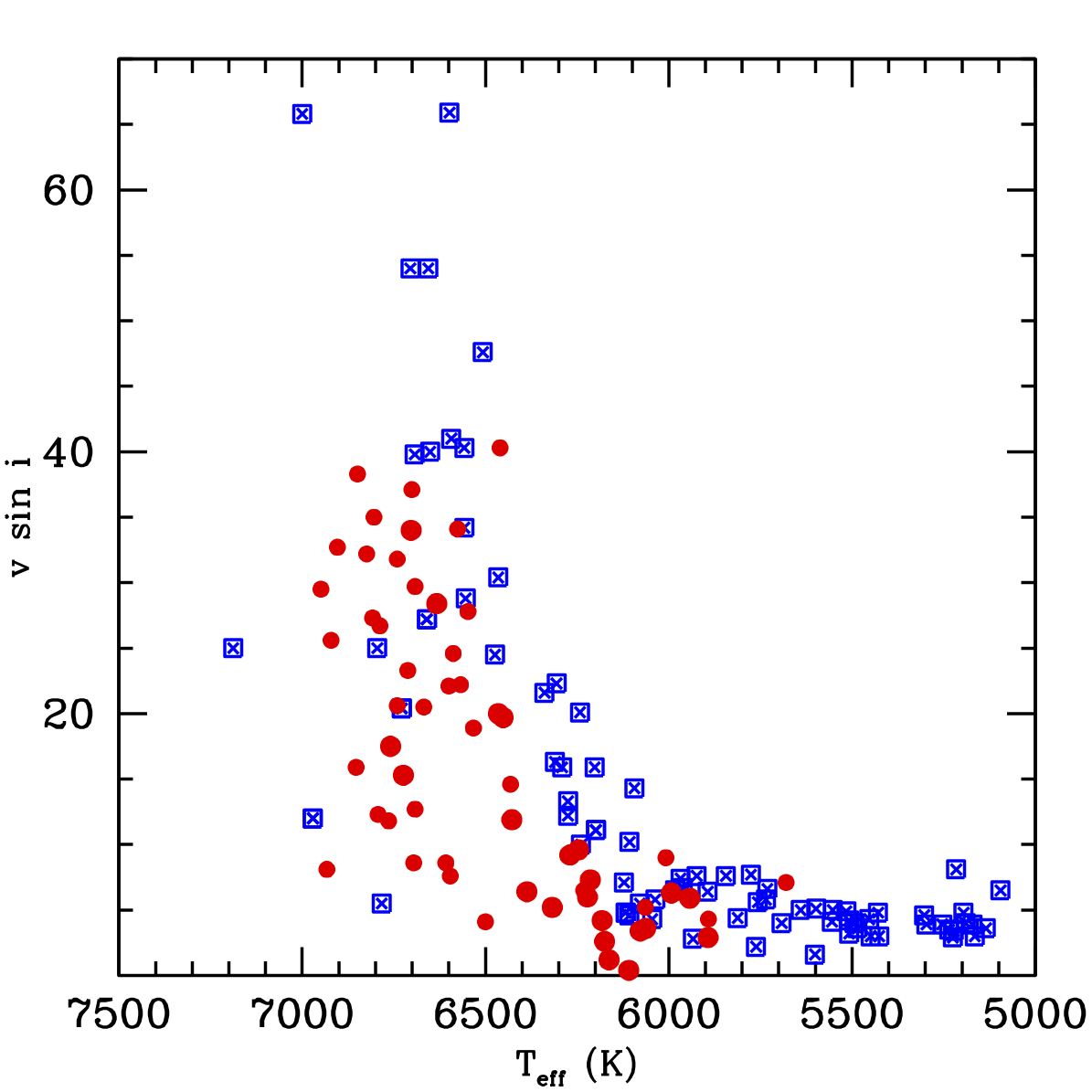}
\caption{Rotational velocities as v sin i with temperature.  The Hyades stars
are blue crossed squares and the NGC 752 stars are red circles with the
smaller circles having temperatures from B-V values.  At a given temperature
the points for NGC 752 are generally below the upper envelope for the Hyades
points.}
\label{fig:f11}
\end{figure}

Their calculations are relevant for our clusters and were done for solar metallicity. Figure \ref{fig:f10} shows our Be abundances in NGC 752 along with the calculated abundances for initial velocities of 10 and 30 km s$^{-1}$ for the younger two ages. This shows that it would not be unexpected for NGC 752, at an age of 1.34 Gyr \citep{agueros2018}, to show some Be depletion. The younger Hyades at 0.65 Gyr might have little or no Be depletion caused by extra mixing from rotational spin-down effects. A distribution of initial rotational velocities and so a spread in Be abundances would be expected.

We have assembled data on present-day stellar rotation rates for these two clusters from \citet{mermilliod2009}. These data are in the form of v sin i, the readily measurable parameter. Figure \ref{fig:f11} is a plot of v sin i with temperature over our range of main-sequence stars. At a given temperature the points for NGC 752 are typically below the upper envelope for the Hyades points. If both clusters started out with similar distributions of rotation velocities, it appears that the stars in NGC 752 have spun down more, as would be expected for an older cluster. The rotationally-induced mixing would have acted to reduce both Li and Be in that cluster more than it would in the Hyades, as observed. Stars in both clusters show evidence of the ``Kraft break'' where the surface rotation has slowed down.

\section{Summary}

We have made observations of Be in 14 F and G dwarfs in the open cluster NGC 752 with the Keck I telescope and HIRES. These were added to our Keck/HIRES spectra of Li in 10 stars. Six of our stars were observed for both Li and Be. Both Li and Be abundances were determined by spectral synthesis using the program \texttt{MOOG}.

Our Li results and those of other studies reported in the literature have been put on a common scale for the atmospheric parameters; we now have consistent results for Li in 39 stars. We find a deep and broad Li dip in NGC 752 which extends down to cool temperatures near 6400 K. Lithium abundances reach a cool star peak near 6200 K. They then fall off rapidly toward 5500 K, a signature of the deepening surface convection zone. Our Be abundances results show a deep dip across the Li dip, but no fall-off in the cooler stars. This indicates that convection does not extend deep enough to reach temperatures near 3.6 $\times$ 10$^6$ needed to destroy Be.

Our parameter scale is the same as that of the Hyades cluster so that direct comparisons could be made of the two clusters in their light element content. NGC 752 is twice the age of the Hyades and has 70\% of its metal content. The well-studied Hyades cluster was discovered to have substantial Li depletion in main-sequence stars in the temperature range $\sim$6400-6800 K by \citet{boesgaard1986} and studied more thoroughly since then, most recently by \citet{boesgaard2016} and \citet{cummings2017}. For the stars cooler than the Li-dip in the Hyades, there is a ``Li-peak'' at roughly 6100-6400 K before the Li abundance declines sharply toward 5000 K as can be seen in Figure \ref{fig:f8}.

The abundances of Be in the Hyades was first studied by \citet{boesgaard2002} and later by \citet{boesgaard2016}. Those Hyades Be abundances as a function of temperature can be seen in Figure \ref{fig:f8}. There is a dip in the Be abundances which corresponds to that in the Li abundances, but the Be dip is not nearly as deep as the Li dip. Additionally, there is no decline in the Be abundances at the cooler temperatures as seen for Li.

\section{Conclusions}

In this work we have found intriguing differences in the light element abundances of the two clusters. There is a Li dip in NGC 752 which is wider than the one in the Hyades, extending down to temperatures $\sim$200 K cooler.
 
There is also a Be dip that corresponds to the one in the Hyades. However, the Be dip in NGC 752 appears to be deeper than the Be dip in the Hyades. When these depletions are measured relative to the meteoritic Be abundances, for NGC 752 Be is lower by at least a factor of 20 compared to a factor of 6 for the Hyades.

In the distribution of Li with temperature in the Hyades there is a peak in A(Li) near 6300 K close to 3.2 (see Figure \ref{fig:f8}). This is close to the value on the hot side of the dip. The corresponding peak for NGC 752 (Figure \ref{fig:f6}) is closer to 6200 K and reaches only 2.85, a factor of $\sim$2 lower in Li content. There is a further decline in Li toward cooler stars in both clusters. Lithium declines steadily toward lower temperatures, but there is less Li at a given temperature in NGC 752 as shown in Figure \ref{fig:f9}.

The pattern for the Be abundances in cooler stars is different from that of Li. As seen in Figure \ref{fig:f8} there is no decline in Be in the Hyades for stars between $\sim$5400$-$6300 K with an average value for A(Be) of 1.33 $\pm$0.06. For NGC 752, however, the average Be content in those G stars appears to be uniformly lower than that in the Hyades. The mean value for those eight stars is A(Be) = 1.20 $\pm$0.06. This is a decrease of 75\% compared to the Hyades. Such a difference could be due to age in the sense that there has been a slow decline in Be for cool stars in the older cluster as the calculations show in Figure \ref{fig:f10}. There is also the possibility that the older NGC 752 had lower Be initially.

The cause of the light element depletions has been linked to many physical characteristics in stellar interiors. The most successful has been mixing driven by rotation and spin-down effects. Figure \ref{fig:f11} shows that the stars in NGC 752 are rotating more slowly than their Hyades counterparts. The mixing of surface material into the interior resulting from the rotational spin-down produces the reduction of both Li and Be. The Li and Be abundances differences between the Hyades and NGC 752 could be attributed to age via the influence of rotational mixing.

%
%

\section*{Acknowledgements}

We are grateful to the W. M. Keck Observatory support astronomers for their knowledgeable assistance during our observing runs. We thank C. J. Ma for help with the Li spectra. A.C. acknowledges support from the National Science Foundation through the Graduate Research Fellowship Program (DGE 1842402). C.P.D. acknowledges support from the National Science Foundation through AST-190945.

\facilities{Keck I: HIRES}

\software{\texttt{MOOG} \citep{sneden1973,sobeck2011}}

\vspace{1cm}
\small
\bibliography{main.bib}

\end{document}